# Surveying Solutions to Securing On-Demand Routing Protocols in MANETs

**Nitish Balachandran**
*Department of Computer Science and Information Systems, BITS Pilani*
nitish.balachandran@gmail.com

---------------------------------------------------------------ABSTRACT---------------------------------------------------------------
A Mobile ad hoc Network or MANET is a wireless network of mobile devices that has the ability to self-configure and self-organise and is characterised by an absence of centralised administration and network infrastructure. An appreciable number of routing protocols used in a typical MANET have left the critical aspect of security out of consideration by assuming that all of its constituent nodes are trustworthy and non-malicious. In this paper, we discuss some of the major threats that such networks are vulnerable to, because of these inherently insecure protocols. The focus is specifically on the source-initiated and on-demand routing protocols. Further, solutions and modifications to these protocols that have been proposed over time, enabling them to mitigate the aforementioned threats to some extent, are also analysed.

-------------------------------------------------------------------------------------------------------------------------------------

-------------------------------------------------------------------------------------------------------------------------------------

## 1. INTRODUCTION

**A** mobile ad hoc network (MANET) consists of nodes that communicate with each other via wireless mediums. These nodes also function as routers and forward packets to others. MANETs, by their very nature, do not have any centralised administration nor require any fixed network infrastructure and this makes it quite easy to set up the network quickly.

Routing in such networks is achieved via the use of a number of specific protocols that attempt to counter the challenge posed by dynamic topologies and are widely in use today. Although numerous ad hoc network routing protocols (Dynamic State Routing (DSR) [1], Ad hoc on-demand vector (AODV) [2]) have been proposed, they have generally assumed an environment where all the nodes are co-operative and trustworthy and no security mechanism has been considered [3][23]. We consider only the on-demand routing protocols AODV and DSR in this paper on account of their dynamic nature, better performance with lesser overhead and also their wider acceptance.

## 2. VULNERABILITIES IN MANETs

MANETs suffer from a number of vulnerabilities owing, primarily, to the fact that they depend on their constituent nodes to function effectively. Malicious nodes may interfere with the smooth functioning of the network through quite a few ways, as described below.

### 2.1 Identity Spoofing

Media Access Control (MAC) and Internet Protocol (IP) addresses are frequently used in MANETs to verify and ascertain the identity of its nodes. However, these addresses may be easily spoofed using tools that are publicly available, which leads to a spoofing attack [5]. In this attack, the malicious user attempts to acquire the identity of a legitimate node in the network. Masquerading as a legitimate user allows the malicious node to avail of privileged services, that are otherwise accessible to only genuine nodes, and become an authorized entity in the network. This attack aims to establish a connection that will enable the attacker to access the sensitive data of the other hosts [6] [7].

## 2.2 Denial of Service (DoS)

DoS is one of the most well known attacks on computer networks largely because of the impact it has on the smooth functioning of the network. This kind of attack is especially damaging to MANETs owing to the limited communication bandwidth and resources of the nodes [8]. In the AODV protocol for instance, a large number of RREQs (message requests) are sent to a destination node on the network that is non-existent. As there is no reply to these RREQs, they will flood the entire network leading to a consumption of all of the node battery power, along with network bandwidth and this could lead to denial-of-service.

## 2.3 Black Hole Attack

The goal of the malicious node in this attack, is to drop all packets that are directed to it instead of forwarding them as intended. It uses its routing protocol in order to advertise itself as having the shortest route to the target node or to any packet that it wants to intercept.

The malicious node advertises its availability of new routes without checking its routing table [12]. In this way the malicious node will always have availability of routes while replying to the route request and hence intercept the data packet. As a result of the dropped packets, the amount of retransmission consequently increases leading to congestion.

There is a more subtle form of this attack wherein the attacker selectively forwards packets instead of dropping all of them altogether. Packets originating from some particular nodes may be modified or suppressed while leaving the data from the other nodes unaffected, thus limiting the suspicion of its malicious behaviour by the other nodes [13].

## 2.4 Wormhole Attack

A wormhole [10] is a combination of two or more malicious nodes belonging to the ad hoc network that are connected by a faster, direct connection. In this kind of attack, an attacker records information at a source or origin, tunnels it through this channel to a target point and retransmits the information in the neighbourhood of the destination. The two nodes are more than one hop away from each other on the network.

Since a wormhole attack can be launched without compromising any node or even the integrity and authenticity of the communication and is implemented with very few resources, it is difficult to detect [11].

## 2.5 Routing Table / Cache Poisoning

In this kind of attack, hostile nodes in the networks send fictitious routing updates or modify genuine route update packets that are sent to other uncompromised nodes. Routing table poisoning may result in sub-optimal routing, congestion in portions of the network, or even make some parts of the network inaccessible [13]. In the case of on-demand routing protocols that we are considering, each node maintains a route cache, that contains information about specific routes that have become well known or have been frequently used and an adversary may also poison the route cache to achieve similar objectives.

## 2.6 Colluding Attacks

This kind of threat is encountered when two or more nodes collude in order to disrupt the smooth functioning of the network by modifying or dropping all packets addressed to them.

Apart from being difficult to detect, it can also cause a significant amount of damage and poses a more serious threat than similar attacks such as the black hole attack. The authors of [14] discuss this in the OSLR protocol and show that two colluding nodes can take down a maximum of 100 percent of data packets.

## 2.7 Rushing Attacks

In [18], the authors discuss an attack called the "Rushing" attack that acts as an effective denial-of-service attack. In this attack, the route request (RREQ) packet sent by the source node to the malicious node is flooded throughout the network by this malicious node quickly enough to prevent other nodes from reacting to the same RREQ. The other nodes that receive the duplicate RREQ from the attacker simply ignore them. Hence, any route discovered by the source node will have the malicious node as an intermediate point in the route [13]. Most of the current on-demand ad hoc routing protocols are vulnerable to this attack due to the fact that most of them use duplicate suppression during the route discovery process. None of the on-demand protocols, including SAODV, SRP, Ariadne, SAR and ARAN (that were all designed to be secure) are affected by rushing attacks since these are extremely difficult to detect.

## 3. METHODS PROPOSED TO SECURE ON-DEMAND ROUTING PROTOCOLS

AODV and DSR are the most popular on-demand protocols. They enable nodes on the network to pass messages, via their neighbours, to nodes with which they cannot directly communicate. This is achieved by discovering routes along which messages can be passed and they ensure that these routes are devoid of loops and try to find the shortest possible path. Further, they are also able to handle changes in routes and can create new routes if there is an error or change in topology, which is a very important requirement for routing in MANETs [4].

Notable mitigation techniques proposed for the vulnerabilities encountered in these on-demand protocols are as under.

### 3.1 Secure AODV (SAODV)

Manel Guerrero Zapata and N. Asokan have proposed SAODV (Secure AODV) in [15], which is an extension of the AODV protocol. Routing messages in SAODV such as route requests and route replies are authenticated to guarantee their integrity and authenticity. The source node signs the routing message with its private key, and the recipient nodes verify the signature using the public key of the source. Since the hop count must be incremented at each hop, the sender is unable to sign it. A hash chain mechanism is used to prevent any modification or tampering of the hop count by hostile intermediate nodes [16].

Fields in the RREQ and RREP packets are authenticated in SAODV. In SAODV, an RREQ packet includes a route request single signature extension (RREQ-SSE). An upper bound for the hop count is chosen by the source node and it generates a one-way hash chain. The length of this one-way hash chain equals the maximum hop count incremented by one. The route request and the anchor of this hash chain are both signed by the source node and both are included in the RREQ single signature extension (RREQ-SSE). Based on the hop count in the route request header, an element of the hash chain is included in the single signature extension.

SAODV however, suffers from performance issues on account of the use of computation-heavy asymmetric cryptography methods. A node is compelled to generate (or verify) a signature each time it generates (or receives) a routing message. A protocol called Adaptive SAODV (A-SAODV) has been proposed by Cerri and Ghioni in [28] to mitigate this.

### 3.2 Authenticated Routing for Ad hoc Networks (ARAN)

ARAN, proposed by Dahill et al. in [20], is a secure routing protocol based on the on-demand protocols. It makes use of asymmetric cryptography and hence a universally trusted digital certification authority (CA) which assigns a digital certificate to every node on the network. ARAN uses peers and trusted third parties to ensure safety in ad hoc networks [19]. The five major components in ARAN are, Certification, Authenticated Route Discovery, Authenticated Route Set up, Route Maintenance, and Key Revocation [29]. All nodes that want to enter the network must request a certificate from the CA.

The initiator node starts the communication by sending the route discovery packet and the target node replies with an RREP packet to the initiator, where it is verified. Every node on any particular route that receives a route request (RREQ), strips the signature and certificate of the previous node and appends its own into the packet before despatching it.

Since the RREQs and RREPs are authenticated for every hop in ARAN, impersonation is easily avoided. As with SAODV, ARAN uses asymmetric key cryptography which slows down the overall performance of the ad hoc network. ARAN is affected by the Wormhole, black hole, DoS attacks [19], relay attacks [22] and also the rushing attack (that all on-demand protocols are vulnerable to) [18].

### 3.3 Security-Aware Ad hoc Routing (SAR)

SAR [30] attempts to implement a more generalized way of providing security to routing protocols. It does this by incorporating security metrics into its base routing protocols (AODV and DSR) themselves (into the RREQ packets) and alters its forwarding behaviour.
When intermediate nodes on a route receive a route request packet having a certain security metric or trust level, they will only be able to use or forward the packet if they can provide the requisite security / trust level.
SAR lets us use security as a metric that in turn, lets the ad hoc routing protocols discover routes efficiently. Though, it is unable to clearly state how to use it as such [31].
SAR prevents a few attacks such as spoofing and the black hole attack among others. However it is found to be vulnerable to DoS, Wormholes and Rushing [18].

### 3.4 Secure Routing Protocol (SRP)

The Secure routing protocol developed by Papadimitratos and Haas in [17] is again an extension to the on-demand protocols. There are three important assumptions made by the authors viz. bidirectional communication, presence of a security association between the source and destination and the absence of colluding nodes. Further, the two communicating nodes must have a shared key for communication and verification [21].

SRP was claimed to guarantee the acquisition of accurate topological information and also that a node initiating the route discovery process would be able to discard replies from malicious nodes claiming false topological information, thus ensuring maximum safety.
Although it prevents the black hole attack and also the attacks due to incorrect routing information, it does not solve the problem of protecting transmitted data (handled by the Secure Message Transmission Protocol) and is only concerned with obtaining correct topological information [29]. It is vulnerable to the wormhole, rushing, DoS and invisible node attacks [22].

### 3.5 Ariadne

Ariadne is a robust protocol based on Dynamic Source Routing that has been proposed by Hu et al. [24] that makes use of the TESLA [25] broadcast authentication technology (which uses the message authentication code (MAC) to verify a message and also has anti-spoofing mechanisms). Ariadne makes use of symmetric key cryptography. It also uses a one way hash along with a MAC using a shared key between the source and the destination in order to authenticate the source at the destination. Every intermediate node on a particular route adds, along with its address, its own message authentication code. As a result, the source node can authenticate all individual entries in the route reply path.

The basic operation of the protocol can be summarised as follows:
- A route request packet is sent out by the initiator when communication is to be commenced. The RREQ has information such as an identifier for the particular route that has been discovered along with a TESLA time interval.
- Upon receipt of the RREQ, the recipient intermediate node checks whether the TESLA time interval is still valid.
- The hash function described earlier is used to check the authentication. Each hop on the path is verified by the target node by comparing the computed hash and the received hash [26].

Ariadne has been shown to be vulnerable to the invisible node attack [27] (where a node participates in a routing protocol that implements identification, without revealing its identity). It is also affected by rushing [18], wormhole, DoS and the invisible node attack as discussed in [22].

### 3.6 Miscellaneous

Apart from secure protocols themselves, a number of individual methods specific to certain kinds of attacks have been suggested to mitigate threats that other protocols may or may not deal with.

For countering black hole attacks, the authors of [33] have proposed a three layer enforcement strategy to secure MANETs by integrating three layers, the prevention layer (based on cryptographic techniques), detection-reaction layer (based on monitoring technique) and enforcement layer (based on obligations). An alternate method to detect single black hole nodes has been discussed in [9].

In case of the DoS or distributed DoS attacks due to route request flooding in MANETs, a scheme based on the addition of the parameters RREQ_ACCEPT_LIMIT and RREQ_BLACKLIST_LIMIT to the original AODV has been proposed in [32].

In the case of Wormhole attacks that affect all of the secure on-demand protocols discussed thus far, Lazos et al. have presented a cryptography-based solution in [11] and have analytically shown that an appropriate choice of network parameters can almost completely eliminate the probability of wormhole links. Apart from this, TIK, a protocol proposed in [34], has been used by the authors to implement temporal packet leashes which defend against the wormhole attack.

In [35], Dey et al. have introduced an improved data hiding technique using prime numbers, which is an improvement over the Fibonacci LSB data hiding method described by Battisti et al. in [36] and this enables secret messages to be embedded in higher bit planes. This idea may be extended to increasing security in Ad hoc networks as well.

A method for detecting the sinkhole attack is proposed in [37] based on an incremental learning algorithm. The authors have confirmed via experiments that this method can be used to detect different kinds of sinkhole attacks as well.

In [18], the authors have introduced Rushing Attack Prevention (RAP) protocol to prevent the powerful rushing attack in routing protocols that use duplicate suppression, such as all of the current on-demand protocols discussed in this paper. RAP also does not incur any extra cost to the underlying routing protocol it is merged with unless the underlying protocol cannot find any valid routes.

Table I: Various on-demand Secure Protocols mapped to the attacks that they are vulnerable to along with their base protocols

| S. No. | Secure on-demand routing protocol | Attacks that the protocols are vulnerable to | Base Protocol |
|---|---|---|---|
| 1 | SAODV | Rushing [18], DoS, Wormhole | AODV |
| 2 | ARAN | Wormhole, Black hole, DoS [19], Rushing [18], and Relay Attacks [22], Wormhole | AODV |
| 3 | SAR | Rushing [18], DoS, Wormhole | AODV / DSR |
| 4 | SRP | Rushing [18], Invisible Node attacks [22], Wormhole, DoS | DSR |
| 5 | Ariadne | Rushing [18], Invisible Node attack [22], Wormhole, DoS | DSR |

## 4. CONCLUSION

In this paper, we have discussed some of the important threats that mobile ad hoc networks using on-demand routing protocols have to encounter. We have also listed some of the notable solutions that have been proposed to provide for security in such MANETs. Further, we have described their modus operandi, key advantages and the threats that they are vulnerable to. TABLE I at the end of section three summarizes the attacks that these secure protocols do not guard against along with the underlying on-demand routing protocol that they are based on.

## REFERENCES

[1] D.B. Johnson and D.A. Maltz, Dynamic Source Routing in Ad Hoc Wireless Networks, *Mobile Computing*, chapter 5, pp. 153-181, (Kluwer Academic Publishers, 1996)

[2] C. E. Perkins, E. M. Belding-Royer, and S. R. Das, Ad hoc On-Demand Distance Vector (AODV) Routing, RFC 3561, Jul 2003.

[3] Du, Xinjun, Ge, Jianhua, Wang,Ying, A method for security enhancements in AODV protocol, *In Proceedings of the17th International Conference on Advanced Information Networking and Applications (AINA'03)*,IEEE Computer Society Washington, DC 2003, pp. 237-240.

[4] Luke Klein-Berndt *A Quick Guide to AODV Rout ing*, National Institute of Standards and Techhnology, US Department of Commerce, USA. Available on: http://w3.antd.nist.gov/wctg/aodv_kernel/aodv_guide.pdf/

[5] Joshua Wright, GCIH, CCNA, 2003, *Detecting wireless LAN MAC addresses spoofing*, technical document. [Online] Available: http://home.jwu.edu/jwright/papers/wlan-mac-spoof.pdf/

[6] Gayathri Chandrasekaran, John-Austen Deymious, Vinod Ganapathy, Marco Gruteser, Wade Trappe, 2009, Detecting Identity Spoofs in 802.11e Wireless Networks, In *Proceeding of the28th IEEE conference on Global telecommunications*, pp.4244-4147

[7] Latha Tamilselvan and Dr. V. Sankaranarayanan, 2007, Prevention of Impersonation Attack in Wireless Mobile Ad hoc Networks, *International Journal of Computer Science and Network Security, Vol.7, No.3* pp.118–123 (2007)

[8] Xiaoxin Wu, David,K., Y.Yau, Mitigating Denial-of-Service Attacks in MANET by Distributed Packet Filtering: A Game theoretic Approach, in *Proceedings of the 2nd ACM symposium on Information, computer and communication security*, pp 365-367 (2006)

[9] Hongmei Deng, Wei Li, and Dharma P. Agarwal Routing security in wireless Ad Hoc networks, *IEEE Communications Magazine, vol. 40, no. 10* October 2002 pp. 70-75

[10] Y. C. Hu, D. B. Johnson, and A. Perrig, SEAD: Secure efficient distance vector routing for mobile , *Elsevier, Vol. 1, Issue 1,* July 2003, Pages 175–192.

[11] L. Lazos, R. Poovendran, C. Meadows, P. Syverson, L.W. Chang, Preventing Wormhole Attacks on Wireless Ad Hoc Networks: A Graph Theoretic Approach, In *Proceedings of Wireless Communications and Networking Conference, 2005*, IEEE March 2005. pp.1193-1199.

[12] G. A. Pegueno and J. R. Rivera, *Extension to MAC 802.11 for performance Improvement in MANET*, Masters Thesis at Karlstads University, Sweden, December 2006

[13] Pradip M. Jawandhiya, Mangesh Ghonge, M.S. Ali and J.S. Deshpande, A Survey of Mobile Ad Hoc Network Attacks, *International Journal of Engineering Science and Technology Vol. 2(9)*, 2010, pp. 4063-4071

[14] B. Kannhavong, H. Nakayama, A. Jamalipour, A Collusion Attack Against OLSR-Based Mobile Ad Hoc Networks, *IEEE GLOBECOM '06,* San Francisco, CA, Nov. 27 2006-Dec. 1, 2006, pp. 1-5.

[15] M. Guerrero Zapata and N. Asokan, Securing Ad hoc Routing Protocols, in *Proceedings of the 1st ACM workshop on Wireless security*, Atlanta, GA, USA, Sep 2002, pp. 1–10

[16] D. Cerri, A. Ghioni, SecuringAODV: The A-SAODV Secure Routing Prototype, *IEEE Communication Magazine*, Feb. 2008, pp 120-125

[17] P. Papadimitratos, and Z. Haas, Secure routing for mobile ad hoc networks, in *Proceedings of the SCS communication Networks and Distributed Systems Modeling and Simulation Conference*, San Antonio, TX, Jan. 27-31,2002, pp. 1-7

[18] Yih-Chun Hu, Adrian Perrig, and David B. Johnson, Rushing attacks and defense in wireless ad hoc network routing protocols, In *Proceeding of the ACM workshop on WIreless SEcurity WISE 2003*, San Diego, CA, USA, September 19, 2003, pp. 1-11

[19] Seema Mehla, Bhawna Gupta and Preeti Nagrath Analyzing security of Authenticated Routing Protocol (ARAN) (IJCSE) *International Journal on Computer Science and Engineering Vol. 02, No. 03*, 2010, pp. 664-668

[20] K. Sanzgiri, B. Dahill, B.N. Levine, E.M. Belding-Royer, A secure routing protocol for ad hoc networks, in *Proceedings of the 10th IEEE International Conference on Network Protocols (ICNP)*, Paris, France, p.78-89, November 12-15, 2002 November 2002.

[21] Yi P, Wu Y, Zou F, Liu N, A Survey on Security in Wireless Mesh Networks, *IETE Tech Rev 2010*; 27: pp. 6-14

[22] Marshall, J, An Analysis of the Secure Routing Protocol for mobile ad hoc network route discovery: using intuitive reasoning and formal verification to identify flaws, Master's Thesis, Department of Computer Science. Florida State University, Tallahassee, FL FSU CS Technical Report #030502, May 2003

[23] Anand Patwardhan, Jim Parker, Anupam Joshi, Michaela Iorga and Tom Karygiannis, Secure Routing and Intrusion Detection in Ad Hoc

Networks. *Proceedings of the 3rd International Conference on Pervasive Computing and Communications (PerCom 2005)* pages 8–12, 2005

[24] Y.C. Hu, P. Adrian, and B. David. Ariadne: A secure on-demand routing protocol for ad hoc networks, in *Proceedings of the MobiCom 2002, Atlanta, Georgia*, USA, Sep. 23-28 2002.

[25] A. Perrig, R. Canetti, D. Song, and J.D. Tygar. Efficient and secure source authentication for multicast, in *Proceedings of Network and Distributed System Security symposium*, pp. 35-46, Feb. 2001.

[26] Tomar, P., Suri, P. K., & Soni, M. K. (2010). A comparative study for secure routing in MANET, *International Journal of Computer Applications, Vol. 4(5),* pp. 17-22.

[27] P. Ramachandran and A. Yasinsac, Limitations of On Demand Secure Routing Protocols, *IEEE Information Assurance Workshop 2004*, June 10-11, 2004, pp. 52-59

[28] Davide Cerri and Alessandro Ghioni, Securing AODV: The A-SAODV Secure Routing Prototype, Communications Magazine, *IEEE In Communications Magazine, IEEE, Vol. 46, No. 2.* pp. 120-125, February 2008

[29] B. Hahill, A Secure Protocol for Ad Hoc Networks, *OEEE CNP*, 2002.

[30] S. Yi, P. Naldurg, and R. Kravets, Security-Aware Ad hoc Routing for Wireless Networks, *Proceedings of the 2nd ACM international symposium on Mobile ad hoc Networking & Computing*, Long Beach, CA, USA, 2001, pp. 299-302.

[31] Ertaul, L., Ibrahim, D., Evaluation of Secure Routing Protocols in Mobile Ad Hoc Networks (MANETs), *The 2009 International Conference on Security and Management SAM'09*, July, Las Vegas.

[32] Dhaval Gada, Rajat Gogri, Punit Rathod, Zalak Dedhia, Nirali Mody, Sugata Sanyal, Ajith Abraham, A Distributed Security Scheme for Ad Hoc Networks, *ACM Crossroads, Special Issue on Computer Security. Volume 11, No. 1*, September, 2004, pp. 1-10.

[33] V. Balakrishnan and V. Varadharajan, Packet Drop Attack: A Serious Threat to Operational Mobile Ad hoc Networks, in *Proceedings of the International Conference on Networks and Communication Systems (NCS 2005)*, Krabi, pp. 89-95, April 2005.

[34] H. Yih-Chun and D. Johnson, Wormhole attacks in wireless networks, *IEEE Journal on Selected Areas in Communications (JSAC), vol. 24, no. 2,* Feb. 2006, , pp. 370-380.

[35] Sandipan Dey, Ajith Abraham, Sugata Sanyal, An LSB Data Hiding Technique Using Prime Numbers, Third International Conference on Intelligent Information Hiding and Multimedia Signal Processing, (IIHMSP 2007),Kaohsiung City, Taiwan, *IEEE Computer Society press, USA,vol.2*, Nov. 26-28, 2007, pp.473-476.

[36] F. Battisti, M. Carli, A. Neri, K. Egiaziarian, A Generalized Fibonacci LSB Data Hiding Technique, 3rd *International Conference on Computers and Devices for Communication (CODEC-06),* Institute of Radio Physics and Electronics, University of Calcutta, December 18-20, 2006

[37] Kim, Kisung and Kim, Sehun, A Sinkhole Detection Method based on Incremental Learning in Wireless Ad Hoc Networks, Department of Industiral Engineering, Korea Advanced Institute of Science and Technology

**Author's Biography**

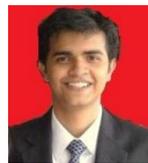

Nitish Balachandran is a final year student of Information Systems at the Birla Institute of Technology & Science (BITS), Pilani, India. His research interests include security in self-organising networks, wireless networks, the TCP/IP suite and privacy protection. He intends to pursue further studies in Networking and Security.